\documentclass[a4paper]{article}

\usepackage{INTERSPEECH2019,url, hyperref}

\title{A non-causal FFTNet architecture for speech enhancement}
\name{Muhammed PV Shifas, Nagaraj Adiga, Vassilis Tsiaras, Yannis Stylianou}
\address{
  Speech Signal Processing Lab (SSPL), University of Crete, Greece}
\email{\{shifaspv,nagaraj,tsiaras,yannis\}@csd.uoc.gr}

\begin{document}

\maketitle

\begin{abstract}
In this paper, we suggest a new parallel, non-causal and shallow waveform domain architecture for speech enhancement based on FFTNet, a neural network for generating high quality audio waveform. In contrast to other waveform based approaches like WaveNet, FFTNet uses an initial wide dilation pattern. Such an architecture better represents the long term correlated structure of speech in the time domain, where noise is usually highly non-correlated, and therefore it is suitable for waveform domain based speech enhancement. To further strengthen this feature of FFTNet, we suggest a non-causal FFTNet architecture, where the present sample in each layer is estimated from the past and future samples of the previous layer. By suggesting a shallow network and applying non-causality within certain limits, the suggested FFTNet for speech enhancement (SE-FFTNet) uses much fewer parameters compared to other neural network based approaches for speech enhancement like WaveNet and SEGAN. Specifically, the suggested network has considerably reduced model parameters: 32\% fewer compared to WaveNet and 87\% fewer compared to SEGAN. Finally, based on subjective and objective metrics, SE-FFTNet outperforms WaveNet in terms of enhanced signal quality, while it provides equally good performance as SEGAN. A Tensorflow implementation of the architecture is provided at ~\footnote{\url{https://github.com/shifaspv/SE-FFTNet-tensorflow-implemenatation}}.
\end{abstract}
\noindent\textbf{Index Terms}: speech enhancement, computational complexity, dilation width, FFTNet, WaveNet

\section{Introduction}
The aim of speech enhancement is to effectively suppress the ambient noise components present in the recorded speech so that to be more intelligible to the listeners. It has application in domains where the background noise suppression is desirable, starting from mobile \& hands-free device users to hearing aids \cite{van2009speech}. The classical speech enhancement methods, like the spectral subtraction \cite{boll1979suppression} and the Wiener filtering \cite{lim1978all}, often rely on the first and second order spectral statistics of the noise. This assumption often fails in real-time applications and leads to a wrong and high variance estimation of the noise statistics which causes severe distortion to the target speech while suppressing background noise. 

To address these limitations neural networks have been widely adopted for speech enhancement task~\cite{xu2014experimental}. This was motivated by the neural architecture ability to extract statistically relevant features using non-linear transformation, starting from the basic convolutional network approach~\cite{park2016fully} to the denoising autoencoder~\cite{lu2013speech} to the more powerful, recurrent neural networks (RNN)~\cite{maas2012recurrent}. RNN based denoising architecture explored the temporal correlation of the speech segments. Hence, it outperforms the former convolutional approach. The most recent approach, further used the long-short-memory cells for the denoising task to store and pass the long term information while doing the prediction~\cite{weninger2015speech}.  All these models explored the non-linearity modeling capability of neural architecture in the \textit{feature domain (magnitude)} of speech, thus ignoring phase information. 

Waveform domain approaches for speech enhancements have been recently suggested: WaveNet~\cite{rethage2018wavenet}, Generative Adversarial Networks (SEGAN)~\cite{pascual2017segan}. These waveform domain models operate on samples of speech by modeling the enhancement task in raw speech waveform. Therefore, they have the potential of using phase information if properly designed.
However, there are some significant limitations of the current waveform domain models: $(1)$ none of these models have given enough attention to the time domain structure of speech and noise while designing their architecture; $(2)$ the computational complexity in terms of model parameters of these models is very high and therefore they are not suitable to implement them in real-time applications.\\
In this work, we suggest a new parallel, non-causal and shallow waveform domain architecture for speech enhancement following a similar convolution pattern as in FFTNet~\cite{jin2018fftnet}. FFTNet has been recently suggested as a fast neural-based audio vocoder. FFTNet makes use of an \textit{initial} wide dilation. Considering an additive noise scenario, such an architecture is very well suited for modeling the long term time-domain correlated structure of target clean speech.
Since we use FFTNet for speech enhancement, we refer to the suggested architecture as Speech Enhancement FFTNet, or SE-FFTNet. In contrast to the original FFTNet auto-regressive structure, SE-FFTNet process the entire input in parallel which significantly increases the prediction speed of the model. Furthermore, SE-FFTNet is a non-causal extension of the original FFTNet. In SE-FFTNet, a shallow architecture is used which has far less number of parameters than other waveform domain methods like SEGAN or WaveNet. Therefore, by combining the parallel and shallow structure, SE-FFTNet has the potential of being applied for real-time applications. The new architecture is trained in end-to-end fashion on a wide range of noise conditions. The results are supported by subjective and objective measures.

The paper is organized as follows. In Section~\ref{sec:theory}, we give more insight into the theory behind the suggested model. In Section~\ref{sec:SE-FFTNet}, the SE-FFTNet architecture is presented. The experimental set-up covering data set and the size of the model is included in Section~\ref{sec:Experimental Setup}. The results are discussed in Section~\ref{sec:results}. Finally, conclusion is given in Section~\ref{sec:conclusion}.

\section{Theoretical Background}
\label{sec:theory}
The neural networks modeling capacity is highly depended on the data set and task on which it is deployed. A model that performs well on images domain may not be the best promising model for speech application, as the speech has rapidly varying samples (16 K samples per second) over time in contrast to the image. This variation should be considered when implementing the neural architecture for speech applications. Even, among the speech applications, differences between the tasks should be taken into account, i.e., the task of Vocoder is very different from that of a speech enhancer. In the specified application of speech enhancement, often the noise in the recorded speech will be less correlated over time than the clean speech. Though many neural models have been suggested for speech enhancement task in recent years, very few of them had given enough attention to the correlation patterns of noisy speech.

In this work, we explored the long-term correlation of speech through an initial wide dilation pattern architecture. In contrast to the traditional waveform models which used the local neighboring samples for extracting the features from the input mixture, the suggested model accounts the wide apart samples of input.  By doing so we expect that the network could effectively discriminate the noise from clean speech. This idea was motivated by the recently proposed FFTNet architecture~\cite{jin2018fftnet}. In FFTNet, the input is split into two equal segments and the merged representation of the two segments is used as input on the next stage. It has been applied successfully in speech synthesis and has a reduced computational complexity compared to other neural-based vocoders. The novelty of this architecture is further important for speech enhancement on exploring the correlation structure of speech and noise.

\begin{figure}[t]
  \centering
  \includegraphics[width=\linewidth, height=5cm]{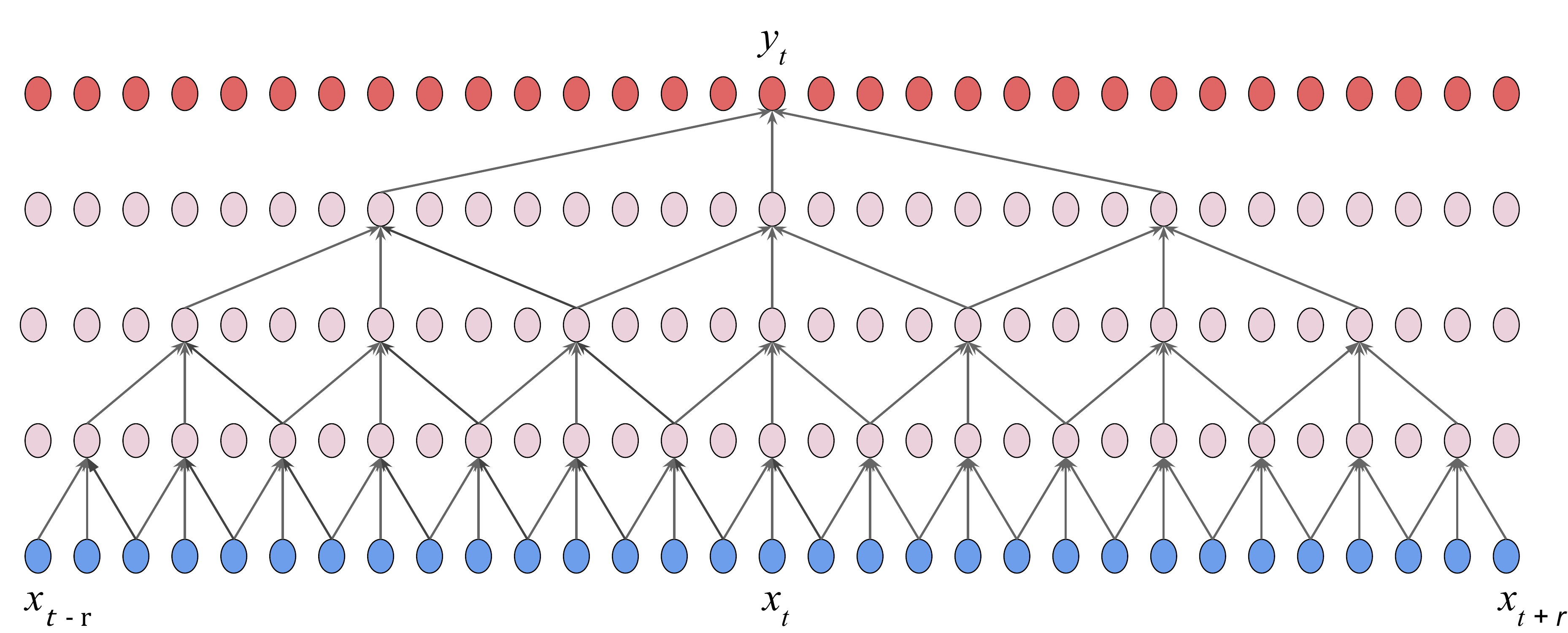}
  \caption{Convolution pattern of SE-WaveNet/ SE-InvFFTNet}
  \label{fig:noncausal_invfftnet}
\end{figure}

\section{Speech Enhancement FFTNet (SE-FFTNet) model}
\label{sec:SE-FFTNet}
The time domain models have the ability to capture high-level acoustic features. Their performance superiority has been proven for many speech applications~\cite{oord2016wavenet}. In the case of speech enhancement, the target is to estimate the clean speech samples from noisy speech samples. As it would be challenging to model the sample distribution of the clean speech from the noisy input, we modeled the denoising task as a regression problem: the model will be looking for the hidden function in the data which represents the mapping from noisy input speech $x_{t}$ to the clean output speech $y_{t}$. This is mathematically formulated in (\ref{eq1}). Here, the objective of the model is to learn the hidden function $f$ from the given data.

\begin{equation}
\hat{y}_t = f(x_{t-r1}, \ldots, x_{t-1}, x_t, x_{t+1}, \ldots, x_{t+r2})
\label{eq1}
\end{equation} 

   The model receptive fields enabled the dependency of past $x_{t-r1}$ and future $x_{t+r2}$ input samples. The model can be causal and non-causal depending on whether to consider the future samples or not, while performing the current sample prediction.    This can be done by controlling the variable $r2$. We have compared the performance of the causal ($r2=0$) and non-causal model ($r2\neq0$) and it has been found that adding non-causality improves the model performance. Hence, in the rest of the paper, the discussion will be on the non-causal model. 
   
\begin{figure}[t]
  \centering
  \includegraphics[width=\linewidth, height=4.9cm]{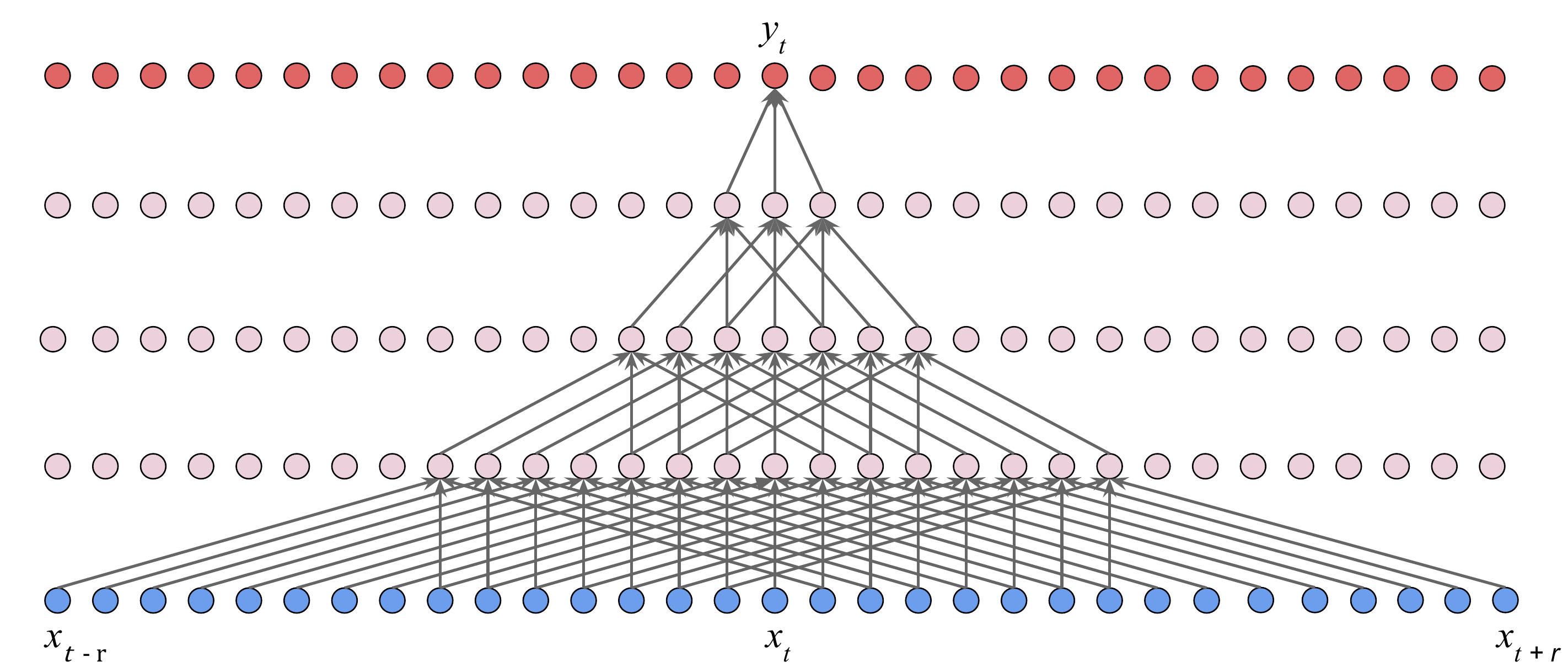}
  \caption{Convolution pattern of the suggested SE-FFTNet}
  \label{fig:noncausal_fftnet}
\end{figure}

In WaveNet~\cite{oord2016wavenet}, sample dependency is introduced by a dilated convolution structure of increasing dilation rate, having the convolution pattern similar to Figure~\ref{fig:noncausal_invfftnet}. This means the first layer of the network extracts the features by looking into the immediate behind and ahead samples. Since the speech and noise variation being equally negligible on these closer time instances, the model may not learn any good discriminating features in its initial layers. This will be rippled on the following layers. To account for this, one must look into the further apart samples of input where time domain correlation for speech is expected, in contrast to noise.  To model this, inspired by FFTNet, the suggested SE-FFTNet have the dilation pattern as shown in Figure~\ref{fig:noncausal_fftnet}. 
We argue that such an architecture will enable SE-FFTNet to easier learn the weights which could discriminate the speech from noise. The similar convolution strategy has been repeated over the layers, until the final enhanced sample is obtained. In other words, SE-FFTNet enables coarser representation at initial layers and finer towards the end. Thus, helping to propagate much cleaner features from the bottom layer to the end.

\begin{figure}[t]
  \centering
  \includegraphics[width=\linewidth]{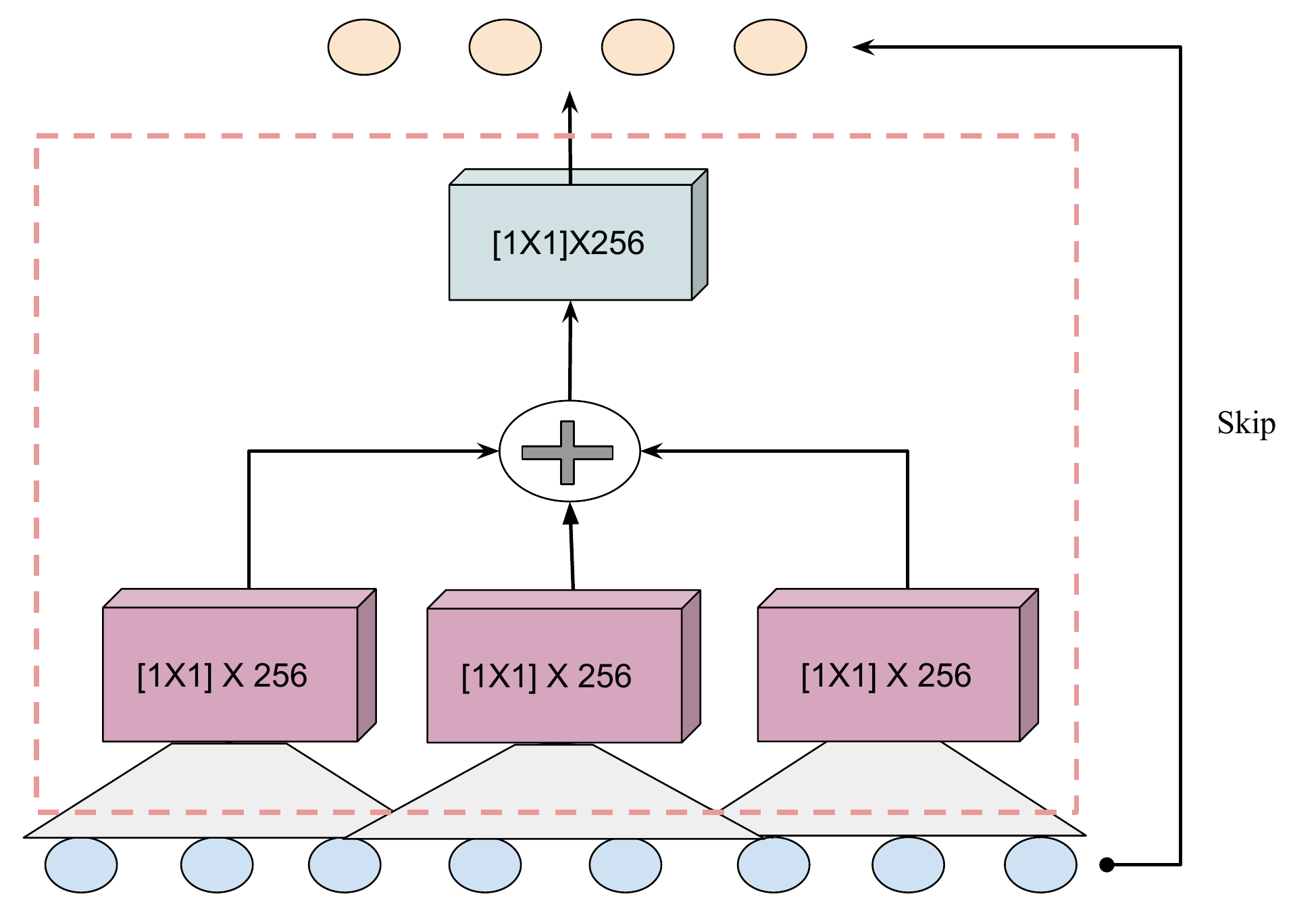}
  \caption{Block insight of SE-FFTNet}
  \label{fig:insight_fftnet}
\end{figure}

In order to evaluate our hypothesis on the influence of initial versus later wide dilation pattern while keeping the internal blocks of the network the same, we suggest to investigate an FFTNet structure where a \textit{later} (similar to WaveNet model) dilation pattern is used. We will refer to that model to as SE-InvFFTNet and that is shown in Figure~\ref{fig:noncausal_invfftnet}. Therefore, the dilation structure of the SE-FFTNet shown in Figure  \ref{fig:noncausal_fftnet} has been inverted so that to have a local neighbouring representation of the input as shown in Figure \ref{fig:noncausal_invfftnet}. It is the dilation pattern similar to the WaveNet presented in SE-WaveNet, but with a difference: The block convolution is retained as shown in Figure \ref{fig:insight_fftnet} in contrast to the WaveNet residual block.  This is needed as the actual WaveNet-SE model and the proposed SE-FFTNet has a different internal block convolution structure connecting to each layer.
     
As the denoising model has to compete with real-time computational constraints we have removed the temporal recurrence on the predicted samples. This means the sample generated at each time instances are totally disjoint, which was not the case in the initial FFTNet model~\cite{jin2018fftnet}. This significantly speeds up the generation process in contrast to the original model while retaining the acoustic modeling ability. The skip connections have been put in place between the layers to facilitate further information flow to the succeeding layer in each level. This is further helpful to restore the phase information which was lost/distorted on passing the signal through the block convolution operations and also, to facilitate gradient back-propagation on training \cite{oord2016pixel}.

The series of operations hidden between the layers are highlighted in Figure~\ref{fig:insight_fftnet}. The past, present and future samples being processed through an one-by-one convolution ($[1\times1]$) of specific channel size.  It is then being sum up into a single representation followed by a ReLU activation. The represenatation then passed through another set of one-by-one convolution ($[1\times1]$) followed by a ReLU activation, to have the final output from the block. This will be added onto the skip bypassing signal from the block input, to have the final input to the next layer. In the end, it is a fully connected layer merges the channel dimension into a speech sample. 

Next we define an appropriate loss function. Since the enhancement task has been formulated as a regression problem, a solution is the mean of the absolute value between the predicted samples and the corresponding clean samples.The distance for the $k$-th training utterance is defined as:
\begin{equation}
L(y^{(k)}, \hat{y}^{(k)}) = \frac{1}{T^{(k)} - 2r} \sum_{t=r}^{T^{(k)}-r}| y^{(k)}_t - \hat{y}^{(k)}_t |
  \label{eq2}
\end{equation}
where, the symbols $y^{(k)}$ and $\hat{y}^{(k)} = \textrm{SE-FFTNet}(x^{(k)})$ correspond to clean signal and to the output of the network, respectively, while $ x^{(k)}$ is the noisy signal. $T^{(k)}$ is the number of samples of the $k$-th utterance and $r$ is the extend of the receptive field. The parameters of the model are tuned in the direction that minimize this loss. The model is trained with noisy speech as input and the corresponding clean speech as the target.

\section{Experimental Setup}
\label{sec:Experimental Setup}

To evaluate the proposed model, 30 speakers were selected from the Voice Bank corpus~\cite{veaux2013voice}. Out of these,  28 speakers were used for training and each speaker data consists of around 400 sentences. To create the noisy mixture, each of these files has been chosen randomly and mixed at a specific SNR point from [0, 5, 10, 15] dB with a selected noise type from the noise set that contains 10 different real-life noises. The different type of noise is selected from DEMAND database\cite{valentini2017noisy}. The remained two speakers were used for testing with the same type of noises used in training but with 4 different SNR level falling in [2.5, 7.5, 12.5, 17.5] dB.  To compare the performance, two recently proposed waveform domain speech enhancement models are considered, namely SEGAN \& SE-WaveNet, which are described below.

\textbf{Speech enhancement GAN (SEGAN):}
Pascual et al.~\cite{pascual2017segan} proposed speech enhancement generative adversarial network (SEGAN). The SEGAN consist of two neural networks, namely, Generator and Discriminator. The Generator network is inspired by Autoencoder architecture. The Generator encoder consists of 11 layers of stride-2 convolution with growing depth, resulting in a feature map at the bottle-neck of 8-time steps with depth 1024. This feature map is concatenated with latent vector "z", sampled randomly from uniform noise distribution.  The resultant concatenated vector is input to an 11-layer up-sampling decoder, with skip connections from corresponding input feature maps. The least square based loss function is used to train SEGAN with additional $L1$ norm to preserve the structure of the enhanced signal. 

\textbf{Speech Enhancement WaveNet (SE-WaveNet):}
Rethage et.al~\cite{rethage2018wavenet} modified the actual WaveNet Vocoder architecture to fit into the speech denoising task. It used a non-causal WaveNet architecture having a dilation pattern similar to Figure \ref{fig:noncausal_invfftnet}, by posing the denoising as a regression task. The model had a series of residual blocks plus the post-processing unit to process the skip outputs from each of these residual blocks. The model was trained to minimize the sample absolute difference objective function, same as in our proposed model. The model had in total 28 residual blocks, with a similar configuration as mentioned in the original paper \cite{rethage2018wavenet}.

\textbf{The model specification:} Both these models have in total 29 layers made up by thrice repeating a block of depth 9 having the dilation factors: [512, 256, 128, 64, 32, 16, 8, 4, 2, 1] for SE-FFTNet and [1, 2, 4, 8, 16, 32, 64, 128, 256, 512] for SE-InvFFTNet. It sums up to a receptive field of size 6138 ( 3069 past \& 3069 future samples), which means it considered 0.38 s of noisy input samples (for 16 kHz signal) when predicting a single clean sample.  In all the layers, one-dimensional convolutions are used with the same number of 256 channels. As the final fully connected layer is being enrolled to merge this channel dimension into a single sample, that has a dimension of [256,1]. During training, the target samples predicted in a single traverse is a set of 4096 (training target field size). The model is fed with a single data point every time with a batch size of 1. In the testing phase, the target field size being varied depends on the test frame length. Just before feeding into the model, the wave files have been normalized to an RMS level of 0.06. This removed the loudness variations among the wave files. The model output loss is minimized with an Adam optimizer of the initial learning rate of 0.001.

As mentioned before, in order to evaluate the influence of dilation steps in the performance, we considered 
and inverted SE-FFTNet architecture as shown in Figure~\ref{fig:noncausal_invfftnet}. We refer this architecture as the SE-InvFFTNet model.

All these models were being trained in a speaker independent fashion. The output speech quality is evaluated both in subjective and objective scale. The perceptual evaluation of speech quality (PESQ) is used as an objective measure of naturalness~\cite{loizou2007speech}. The Short-time objective intelligibility (STOI) score is used to measure the intelligibility gain by processing the noisy mixture, in reference to the clean~\cite{loizou2007speech}. The gain in SNR through the model processing is being evaluated by segmental SNR (SSNR) scale~\cite{loizou2007speech}. The speech distortion and the residual noise intrusion on the enhanced signal are measured with CSIG \& CBAK along with the overall quality of the signal with COVL~\cite{hu2008evaluation}.

The subjective evaluation was done with non-native English listeners listened to the processed samples from different models. To cover the entire test set we have used both lower and high SNR samples while selecting the sentences for listening experiments. They were asked to rate the quality of the samples on a scale of 1-5. In total 15 responses were collected and averaged across all the participants to get the final mean opinion score (MOS).

\section{Results and Discussions}
\label{sec:results}
The models testing is done over 824 files from the test set comprised of different noises. Hence, the results displayed are an average performance on the test set. Table~\ref{tab:Table.1} included the objective performance gain of the proposed SE-FFTNet model along with its competitors. It is clear that the SE-FFTNet model outperforms both the waveform based SE-GAN or SE-WaveNet models. This improvement is reflected in all the subjective metrics in Table~\ref{tab:Table.1}. The higher values in CBAK and CSIG is a clear indication of the model capability to suppress the noise components in the signal without distorting the target speech of interest. The same trend can be seen on the COVL score which is a reflection of the overall signal quality.  This is even more clear when we look into the segmental SNR gain through the processing. SSNR has been increased around 1 dB by processing with SE-FFTNet in comparison to the SE-WaveNet method. 

\newcommand{\ra}[1]{\renewcommand{\arraystretch}{#1}}
\begin{table}[th]
\caption{The Objecive measurements comparing the performance among the models}
 \label{tab:Table.1}
 \centering{
 \ra{1.3}
\scalebox{0.75}{
\begin{tabular}{c c c c c c} 
     \toprule
 {\bf Metric} & {\bf Noisy} & {\bf SEGAN} & {\bf SE-WaveNet} & {\bf SE-FFTNet} & {\bf SE-InvFFTNet}\\ 
      \toprule
 PESQ &1.96 &2.24 &2.23  &{\bf 2.37} &2.24 \\
 STOI &0.28 &0.87 &0.86  &0.87       &0.87 \\ 
 CSIG &3.35 &3.34 &3.33  &{\bf 3.60}  &3.31\\ 
 CBAK &2.44 &3.09 &3.00  & {\bf 3.20} &3.13\\ 
 COVL &2.63 &2.78 &2.76  &{\bf 2.98}  &2.77\\ 
 SSNR &1.63 &9.18 &8.12  &{\bf9.65}   &9.61\\ 
  
 \bottomrule
\end{tabular}}}
\end{table}
The results from the MOS study is displayed in Table~\ref{tab:Table.2}. Though the SE-FFTNet has got higher scores compared to all the other models, the model is slightly under scored compared to the SEGAN.\\
\begin{table}
\caption{MOS with standard error for different methods}
\label{tab:Table.2}
\centering
\scalebox{0.8}{
\begin{tabular}{ c c c c c} 
     \toprule
  Noisy & SEGAN & SE-WaveNet & SE-FFTNet& Inv-FFTNet\\ 
      \toprule
 2.67$\pm$0.12 &3.51$\pm$0.09 &2.8$\pm$0.10 &3.27$\pm$0.10 &2.91$\pm$0.09 \\
  
 \bottomrule
\end{tabular}}
\end{table}

\subsection{Performance between SE-FFTNet \& SE-InvFFTNet} 
The reason behind SE-FFTNet performance improvement might be attributed to the initial hypothesis we have mentioned, where the initial wider dilation of the proposed SE-FFTNet model being enabled a better extraction of the features which could discriminate the noise on the input. By this assumption, the SE-FFTNet should outperform the SE-InvFFTNet. From Table~\ref{tab:Table.1}, all the readings show an inline relation to our assumption. The CSIG gain from 3.31 to 3.60 is a strong sign of target speech restoration by the SE-FFTNet model compared to SE-InvFFTNet. At the same time, noise suppression (CBAK) has been improved from 3.13 to 3.20. Hence the overall quality of the output speech (COVL) is got improved by 0.21.  A similar trend can be observed in the MOS test results displayed in Table~\ref{tab:Table.2}. This is a clear indication that the model with decreasing dilation fields (SE-FFTNet) performs better than the one with increasing dilation (SE-InvFFTNet) for the speech enhancement task. This validated the hypothesis on which the model was built. The enhanced samples from all these models can listen from this link~\footnote{\url{https://www.csd.uoc.gr/~shifaspv/IS2019-demo}}.
\begin{table}[h]
\caption{Total number of model parameters in Million (M)}
\label{tab:Table.3}
\centering
\begin{tabular}{ c c c} 
     \toprule
  SEGAN & SE-WaveNet & SE-FFTNet\\ 
      \toprule
 193 M &34.3 M &23.5 M \\
  
 \bottomrule
\end{tabular}
\end{table}
\subsection{Complexity of the models}

In the real-time application of these neural network based speech enhancement algorithms, the complexity is the biggest constraint. In Table~\ref{tab:Table.3}, we have listed the number of parameters used in SEGAN, SE-WaveNet, and SE-FFTNet models. The complexity displayed is the \textit{testing} complexity of the model. Note that the training of model like SEGAN needs additional parameters for the discriminator network. From Table~\ref{tab:Table.3}, it is clear that the suggested SE-FFTNet model has a far lesser number of parameters compared to others; 32\% lesser than the WaveNet and 87\% lesser than the SEGAN.  This reduction in parameter further highlights the potential of the proposed model towards real-time enhancement applications. One must note that this reduction is accompanied by the performance equal or higher, compared to the existing models.

\section{Conclusions}
\label{sec:conclusion}
In this work, a new parallel, non-causal and shallow waveform domain architecture for speech enhancement based on FFTNet, referred to as SE-FFTNet, is suggested. SE-FFTNet model explored the underlying time domain structure of speech and noise which is important for enhancement. The wider dilation in the initial layers of the model enabled it to learn clean speech structure effectively from the input noisy mixture. The results confirm that the model with a decreasing dilation pattern over depth (SE-FFTNet) performs better than the model with increasing dilation pattern (SE-InvFFTNet). This finding on the influence of dilation width will be useful while implementing the new architecture in future speech enhancement models. The subjective and objective comparative study confirmed the model effectiveness over already existing stat-of-the-art waveform domain models for speech enhancement. In terms of complexity, SE-FFTNet have far less parameters than SE-WaveNet and SEGAN. This reduction in number of parameters of SE-FFTNet shows that it has the potential to be implemented in real-time applications.  Future work includes testing of the proposed model in convolutive noise.

\section{Acknowledgements}

This work was partly funded by the E.U. Horizon2020 GrantAgreement 675324, Marie Sklodowska-Curie Innovative Train-ing Network, ENRICH. 

\bibliographystyle{IEEEtran}
\bibliography{mybib}

\begin{thebibliography}{10}
\providecommand{\url}[1]{#1}
\csname url@samestyle\endcsname
\providecommand{\newblock}{\relax}
\providecommand{\bibinfo}[2]{#2}
\providecommand{\BIBentrySTDinterwordspacing}{\spaceskip=0pt\relax}
\providecommand{\BIBentryALTinterwordstretchfactor}{4}
\providecommand{\BIBentryALTinterwordspacing}{\spaceskip=\fontdimen2\font plus
\BIBentryALTinterwordstretchfactor\fontdimen3\font minus
  \fontdimen4\font\relax}
\providecommand{\BIBforeignlanguage}[2]{{%
\expandafter\ifx\csname l@#1\endcsname\relax
\typeout{** WARNING: IEEEtran.bst: No hyphenation pattern has been}%
\typeout{** loaded for the language `#1'. Using the pattern for}%
\typeout{** the default language instead.}%
\else
\language=\csname l@#1\endcsname
\fi
#2}}
\providecommand{\BIBdecl}{\relax}
\BIBdecl

\bibitem{van2009speech}
T.~Van~den Bogaert, S.~Doclo, J.~Wouters, and M.~Moonen, ``Speech enhancement
  with multichannel wiener filter techniques in multimicrophone binaural
  hearing aids,'' \emph{The Journal of the Acoustical Society of America}, vol.
  125, no.~1, pp. 360--371, 2009.

\bibitem{boll1979suppression}
S.~Boll, ``Suppression of acoustic noise in speech using spectral
  subtraction,'' \emph{IEEE Transactions on acoustics, speech, and signal
  processing}, vol.~27, no.~2, pp. 113--120, 1979.

\bibitem{lim1978all}
J.~Lim and A.~Oppenheim, ``All-pole modeling of degraded speech,'' \emph{IEEE
  Transactions on Acoustics, Speech, and Signal Processing}, vol.~26, no.~3,
  pp. 197--210, 1978.

\bibitem{xu2014experimental}
Y.~Xu, J.~Du, L.-R. Dai, and C.-H. Lee, ``An experimental study on speech
  enhancement based on deep neural networks,'' \emph{IEEE Signal processing
  letters}, vol.~21, no.~1, pp. 65--68, 2014.

\bibitem{park2016fully}
S.~R. Park and J.~Lee, ``A fully convolutional neural network for speech
  enhancement,'' \emph{arXiv preprint arXiv:1609.07132}, 2016.

\bibitem{lu2013speech}
X.~Lu, Y.~Tsao, S.~Matsuda, and C.~Hori, ``Speech enhancement based on deep
  denoising autoencoder.'' in \emph{Interspeech}, 2013, pp. 436--440.

\bibitem{maas2012recurrent}
A.~Maas, Q.~V. Le, T.~M. O’neil, O.~Vinyals, P.~Nguyen, and A.~Y. Ng,
  ``Recurrent neural networks for noise reduction in robust asr,'' 2012.

\bibitem{weninger2015speech}
F.~Weninger, H.~Erdogan, S.~Watanabe, E.~Vincent, J.~Le~Roux, J.~R. Hershey,
  and B.~Schuller, ``Speech enhancement with lstm recurrent neural networks and
  its application to noise-robust asr,'' in \emph{International Conference on
  Latent Variable Analysis and Signal Separation}.\hskip 1em plus 0.5em minus
  0.4em\relax Springer, 2015, pp. 91--99.

\bibitem{rethage2018wavenet}
D.~Rethage, J.~Pons, and X.~Serra, ``A wavenet for speech denoising,'' in
  \emph{2018 IEEE International Conference on Acoustics, Speech and Signal
  Processing (ICASSP)}.\hskip 1em plus 0.5em minus 0.4em\relax IEEE, 2018, pp.
  5069--5073.

\bibitem{pascual2017segan}
S.~Pascual, A.~Bonafonte, and J.~Serr{\`a}, ``Segan: Speech enhancement
  generative adversarial network,'' \emph{arXiv preprint arXiv:1703.09452},
  2017.

\bibitem{jin2018fftnet}
Z.~Jin, A.~Finkelstein, G.~J. Mysore, and J.~Lu, ``Fftnet: A real-time
  speaker-dependent neural vocoder,'' in \emph{2018 IEEE International
  Conference on Acoustics, Speech and Signal Processing (ICASSP)}.\hskip 1em
  plus 0.5em minus 0.4em\relax IEEE, 2018, pp. 2251--2255.

\bibitem{oord2016wavenet}
A.~v.~d. Oord, S.~Dieleman, H.~Zen, K.~Simonyan, O.~Vinyals, A.~Graves,
  N.~Kalchbrenner, A.~Senior, and K.~Kavukcuoglu, ``Wavenet: A generative model
  for raw audio,'' \emph{arXiv preprint arXiv:1609.03499}, 2016.

\bibitem{oord2016pixel}
A.~v.~d. Oord, N.~Kalchbrenner, and K.~Kavukcuoglu, ``Pixel recurrent neural
  networks,'' \emph{arXiv preprint arXiv:1601.06759}, 2016.

\bibitem{veaux2013voice}
C.~Veaux, J.~Yamagishi, and S.~King, ``The voice bank corpus: Design,
  collection and data analysis of a large regional accent speech database,'' in
  \emph{2013 International Conference Oriental COCOSDA held jointly with 2013
  Conference on Asian Spoken Language Research and Evaluation
  (O-COCOSDA/CASLRE)}.\hskip 1em plus 0.5em minus 0.4em\relax IEEE, 2013, pp.
  1--4.

\bibitem{valentini2017noisy}
C.~Valentini-Botinhao \emph{et~al.}, ``Noisy speech database for training
  speech enhancement algorithms and tts models,'' 2017.

\bibitem{loizou2007speech}
P.~C. Loizou, \emph{Speech enhancement: theory and practice}.\hskip 1em plus
  0.5em minus 0.4em\relax CRC press, 2007.

\bibitem{hu2008evaluation}
Y.~Hu and P.~C. Loizou, ``Evaluation of objective quality measures for speech
  enhancement,'' \emph{IEEE Transactions on audio, speech, and language
  processing}, vol.~16, no.~1, pp. 229--238, 2008.

\end{thebibliography}


\end{document}